\documentclass[conference,compsoc]{IEEEtran}

\usepackage{cite}
\usepackage{amsmath,amssymb,amsfonts}
\usepackage{algorithmic}
\usepackage{graphicx}
\usepackage{authblk}

\usepackage{multirow}

\usepackage[table,xcdraw]{xcolor}

\usepackage{textcomp}
\usepackage{xcolor}

\begin{document}
\title{Towards 5G Zero Trusted Air Interface Architecture}
\author{Sheng Sun\qquad Morris Repeta \qquad Mike Healy  \qquad Vish Nandall \qquad  Eddy Fung \qquad Chris Thomas 
\\
Dell Technologies\\
\textit {\{robert.sun, morris\_repeta, mike.healy, vishwamitra.nandall, eddy.fung, chris\_d\_thomas\}@dell.com}}

\maketitle

\begin{abstract}

5G is destined to be supporting large deployment of Industrial IoT (IIoT) with the characteristics of ultra-high densification and low latency. 5G utilizes a more intelligent architecture, with Radio Access Networks (RANs) no longer constrained by base station proximity or proprietary  infrastructure.  The 3rd Generation Partnership Project (3GPP) covers telecommunication technologies including RAN, core transport networks and service capabilities. Open RAN Alliance (O-RAN) aims to define implementation and deployment architectures, focusing on open-source interfaces and functional units to further reduce the cost and complexity. O-RAN based 5G networks could use components from different hardware and software vendors, promoting vendor diversity, interchangeability and 5G supply chain resiliency. Both 3GPP and O-RAN 5G have to manage the security and privacy challenges that arose from the deployment. Many existing research studies have addressed the threats and vulnerabilities within each system.  5G also has the overwhelming challenges in compliance with privacy regulations and requirements which mandate the user identifiable information need to be protected. 

In this paper, we look into the 3GPP and O-RAN 5G security and privacy designs and the identified threats and vulnerabilities.  We also discuss how to extend the Zero Trust Model to provide advanced protection over 5G air interfaces and network components.
\end{abstract}


\section{Introduction}
The 3rd Generation Partnership Project (3GPP), has been actively developing technical specifications for 5G networks, including security specifications. These specifications bring several security enhancements to address new cybersecurity and privacy requirements. 3GPP SA3 is responsible for identifying the security and privacy requirements and defining the security architectures and associated protocols to address these requirements. 3GPP SA3 also ensures that cryptographic algorithms which need to be part of the 5G security specifications are available.

In addition to 3GPP, other standardization bodies and industry groups have been working on developing related  technical specifications and standards. Some of these standards, such as IETF and IEEE, form the basic building blocks of the mechanisms incorporated in 3GPP security specifications. O-RAN architecture has its roots in 3GPP 5G technologies but is also inherently different than 3GPP RAN, due to the goal is to accommodate open interfaces from various vendors/suppliers and lower the cost by migrating functions to the cloud. O-RAN has potentially opened more attacking surfaces due to its openness and cloud deployment.

5G security and privacy architecture is traditionally employing a centralized trust security model, specifically the boundary is established beyond the gNB and the northbound interfaces, any entities and functions inside the perimeter, are assumed to be trusted. Perimeters are usually protected by security measures such as firewalls or intrusion detection systems. This centralized trust model is less effective in coping with cloud computing and virtualization,  as manifested in many attacks on the inside entities and interfaces \cite{b3,b4,b5}

Zero Trust Architecture (ZTA) has emerged as one of the promising design options to protect infrastructure from security threats and vulnerabilities. The essential principle of ZTA is to remove the trust perimeter compared with the centralized based trust model, which has a predetermined border protected by firewalls or Intrusion Detection System/Intrusion Protection System (IDS/IPS).  Many recent studies on cloud-based infrastructure have indicated that inside breaches or attacks are the epicenters of the attacking vectors. On the contrary, ZTA proposes a new security design principle, which is any device, system, user, or application should not be inherently trusted regardless of their location, at any given time, trust shall always be earned and verified.  Embedding ZTA within 5G security and privacy protection design can significantly relieve the concerns from inside attacks. 

In response to the May 2021 “Executive Order on Improving the Nation’s Cybersecurity” \cite{b47} which requires agencies to plan and 
move toward implementing advanced zero trust architectures for the protection of the Federal Government’s information resources, many US agencies had published a series of ZTA strategies, including NIST SP 800-207 "Zero Trust Architecture" \cite{b33}, and US DoD's 5G Strategy and Implementation \cite{b45}, in which the ubiquitous connectivity through 5G networks is perceived by the U.S. DoD as a critical strategic technology that provides nations with long-term economic and military advantage. Next-generation networks are especially important for mission-critical communications and tactical edge net-works (TEN). However, perimeter-based security models exhibit weaknesses in providing network assurance in a heterogeneous and dynamic network environment. Further, the operation of intelligent TEN might heavily rely on cloud-based services for data management and processing. Hence, the  Zero Trust Model  for such highly mobile networks is necessary for providing information security of 5G infrastructure. 

The paper is structured as follows, we briefly introduce the security and privacy security goals of 3GPP and O-RAN respectively in section 1; in section 2, we deep dive into the details of 3GPP 5G security and privacy design over its architecture, at end of the section, we also concluded the security threats and vulnerabilities; in section 3, we look into the details of O-RAN security architecture and its protection design over the open interfaces, a summary of potential security threats and vulnerabilities is provided at the end of the chapter; in section 4, we discuss the Zero Trust Model and how it could be applied into the design of O-RAN security architecture, and also we will analyze the cost efficiency challenges when applying the zero trust model to 5G; in chapter 5, we will conclude the paper with future plans.

\section{3GPP 5G security}

In 3GPP TS 33.501 specification \cite{b6}, it outlines all the security capabilities and functions within 5G architecture. A high-level 5G architecture comprises 7 security domains within 3 stratum: Application Stratum, Home Stratum/Serving Stratum and Transport Stratum, from the top layer to the bottom layer respectively, as shown in figure 1.  
\begin{figure}[ht]
    \centering
    \includegraphics[width=1\columnwidth]{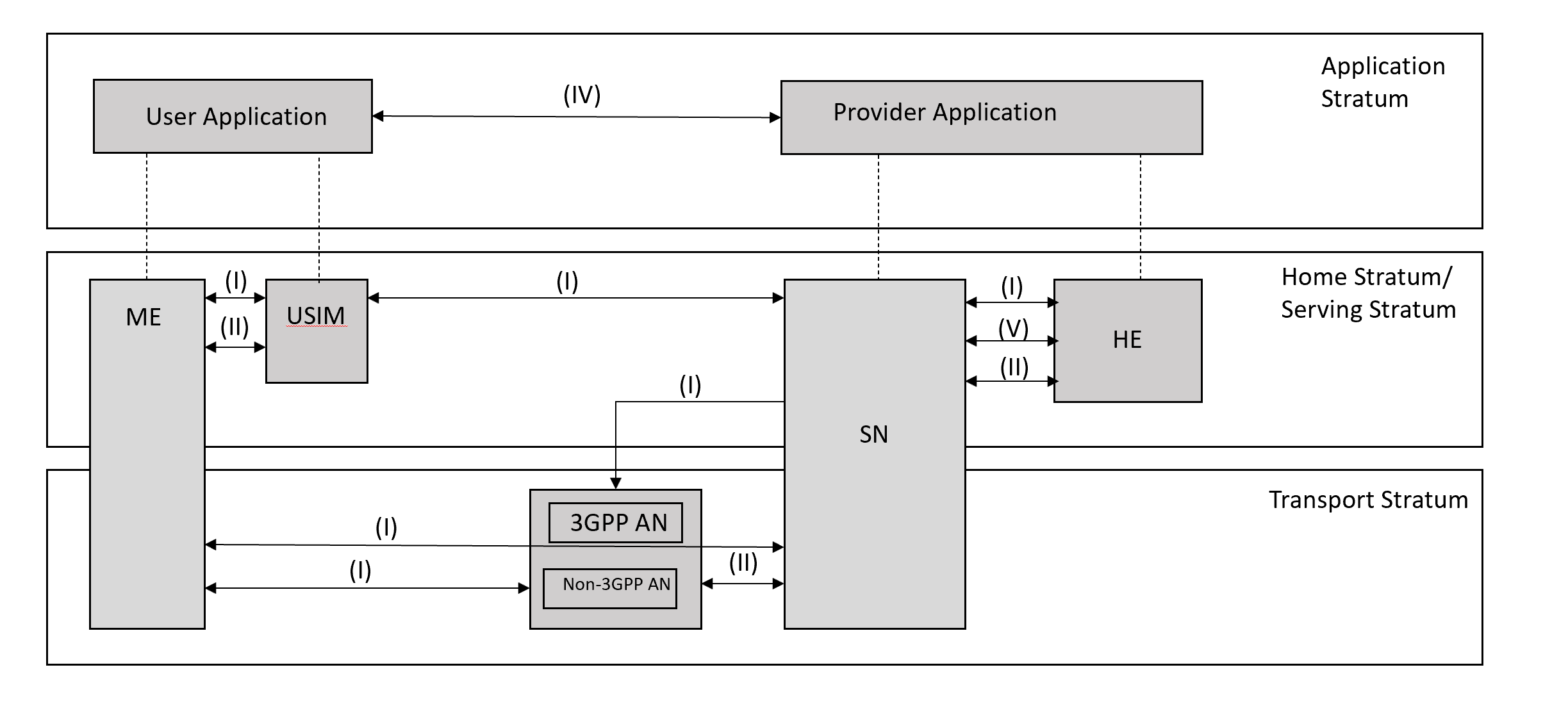}\hspace{2pc} 
 
    \caption{5G High Level Security Architecture}
    \label{fig:my_label}
\end{figure}

\begin{itemize}
    \item Network access security (I): includes the set of security features that enable a UE to authenticate and access services via 
the network securely, including the 3GPP access and Non-3GPP access, and in particularly, to protect against 
attacks on the (radio) interfaces. In addition, it includes the security context delivery from SN to AN for the 
access security. 
    \item Network domain security (II): includes the set of security features that enable network nodes to securely exchange 
signalling data and user plane data. 
    \item User domain security (III): includes the set of security features that secure the user access to mobile equipment. 
    \item Application domain security (IV): includes the set of security features that enable applications in the user domain and in 
the provider domain to exchange messages securely. Application domain security is out of scope of the present 
document. 
    \item SBA domain security (V):includes the set of security features that enable network functions of the SBA architecture to securely communicate within the serving network domain and with other network domains. Such features include network function registration, discovery, and authorization security aspects, as well as the protection for the service-based interfaces. SBA domain security is a new security feature compared to TS 33.401 
\cite{b9}.
    \item Visibility and configurability of security (VI): includes the set of features that enable the user to be informed whether a 
security feature is in operation or not.
\end{itemize}

5G specifications bring significant security improvements in comparison to previous generations of networks. In the  following sections we will highlight the security features. 

\subsection{Authentication Framework}
5G-AKA is one of the techniques available in 5G for mutual authentication between the subscriber and the network, as well as a key agreement for the protection of NAS, RRC and User Plane traffic. In 5G, EAP-AKA is also supported to enhance the roaming and N3WI interworking security. 

5G also defines the  authentication mechanisms for network slicing in TS 33.501, in which Clause 16.3 defines requirements for network slice-specific authentication and  authorization  between  UE and external AAA servers. These requirements specify that the EAP framework is to  be used for this purpose with SEAF/AMF as the EAP authenticator.

\subsubsection{AKA authentication with concealed identity}
In 5G, one of the significant security features is the subscriber privacy protection. A few earlier studies have exposed some of the attacks which resulted in the IMSI breach \cite{b13} caught by a false base station. The root reason is due to the design vulnerability of IMSI ( international mobile subscriber identity), which was sent in clear text over the air. In 5G, the authentication has replaced IMSI with concealed subscriber identifiers, such as SUCI and GUTI. 
\begin{figure}[ht]
    \centering
    \includegraphics[width=1\columnwidth]{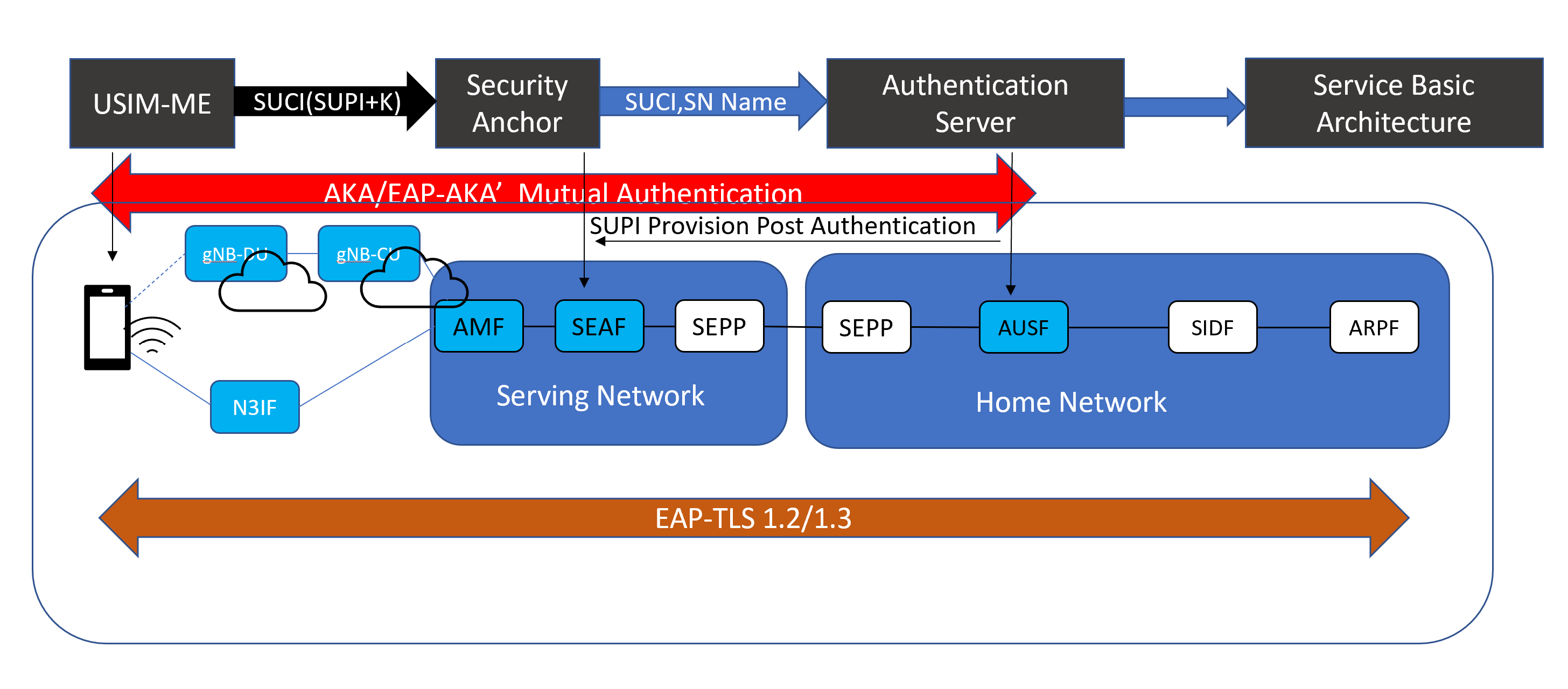}\hspace{2pc}
 
    \caption{5G security and identity management}
    \label{fig:my_label}
\end{figure}

When the 5G AKA authentication protocol starts the network access, the UE presents the concealed ID with SUCI or GUTI to perform authentication with the AUSF (Authentication Server Function). It uses services of UDM (Unified Data Management) and ARPF (Authentication Credential 
Repository and Processing Function), during the authentication handshake between the UE and AUSF, SIDF (Subscriber Identifier De-concealing Function) comes into play to recover SUPI from SUCI.

Once the subscriber identity and keying material are authenticated, the AKA protocol will derive the keying materials for a set of functions, including, confidentiality key (CK), Integrity key (IK) and roaming keying materials. In 5G, SEAF provides the proxy between serving network with the home networks, it also holds the intermediate key $K_{SEAF}$. Keys for more than one security context can then be derived 
from the $K_{SEAF}$ without the need of a new authentication run. For example, keys obtained from an authentication run  over a 3GPP access network can be used to establish security between the user equipment and a Non-3GPP access InterWorking Function (N3IWF) used in untrusted non-3GPP access.

\begin{figure}[ht]
    \centering
    \includegraphics[width=1\columnwidth]{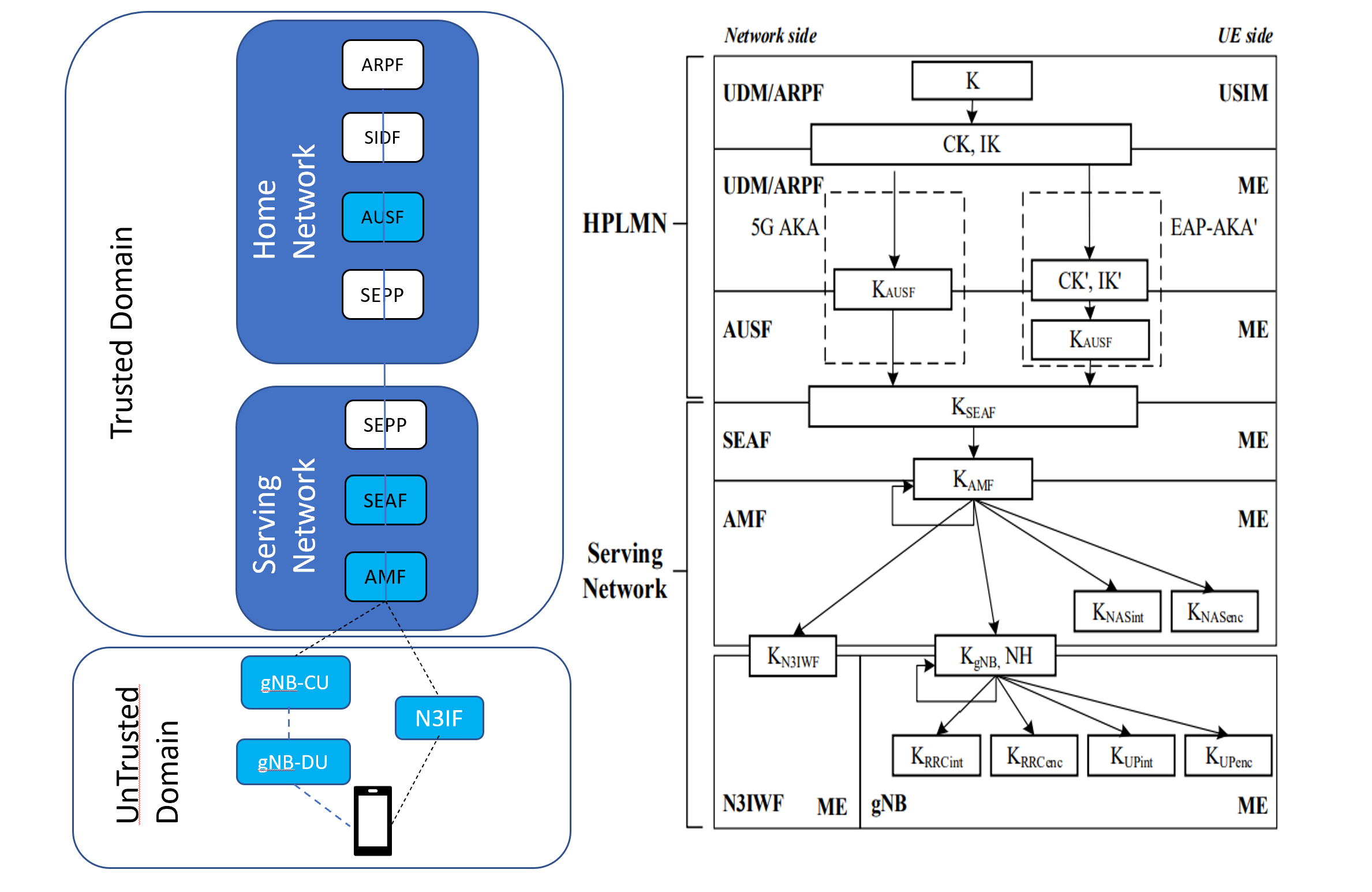}\hspace{2pc}
 
    \caption{5G Key Hierarchy and Generation}
    \label{fig:my_label}
\end{figure}

\subsubsection{EAP-AKA authentication}
5G networks also optionally support the EAP-AKA (Extensible Authentication Protocol) defined in \cite{b14,b15}, EAP-AKA provides flexibility for 
authenticating 3GPP and non-3GPP access networks, such as WLAN.

\subsection{Subscriber's Privacy Protection}
In 3GPP TS33.501, 5G AKA and EAP-AKA provides a mechanism to conceal the subscriber's identity during the authentication handshake, which prior to 5G and earlier, the IMSI was transmitted in clear text and resulted in some attacks. Substantial effort in addressing the privacy exploits by tools, such as  International Mobile Subscriber Identifier (IMSI) catchers or Stingrays \cite{b10} have been investigated and specified in 5G. Authentication mechanism employs the use of permanent, concealed and temporary subscription identifiers. The permanent identifier is referred to as Subscription Permanent Identifier (SUPI) which is globally unique ID to replace the IMSI in 4G and earlier versions. The concealed ID is called Subscription Concealed Identifier (SUCI) which is cryptographically generated at UE.

The subscription temporary identifier is called Globally Unique Temporary UE Identity (GUTI) which is sent to the user-equipment after a successful activation of non-access stratum security.

\subsection{Roaming Security}
Security issues in roaming and inter-operator interfaces have been utilized for attacking and the fraudulent access in 4G, due to the old version of signalling (S7 and SIGTRAN), and the AAA service in the core. 5G has developed new schemes to enhance the roaming security. In TS 33.501,  the Security Edge Protection Proxy (SEPP)  acts as the security gateway on interconnections between the home network and visited networks. SEPP supports the TLS 1.2 to provide the authentication, and end-to-end encryption
Functions supported by SEPPs include end-to-end authentication, integrity and confidentiality protection. 

SEPP uses JSON Web Encryption described in IETF RFC 7516 (JSON Web Encryption (JWE)) for protecting messages on 
the N32 interface against eavesdropping and replay attacks and IP exchange service providers use JSON Web 
Signatures defined in IETF RFC 7515 (JSON Web Signature (JWS)) for signing their modifications needed for their 
mediation services. The application layer security protocol for the N32 interface is called PRINS.

5G roaming architecture also supports the  OAuth 2.0 authorization framework \cite{b16}  in the context of authorization for Network Function (NF) service access within 
the Public Land Mobile network (PLMN) and roaming scenarios.

\subsection{Cryptographic Modules and primitives}

\subsubsection{Encryption algorithms}

Four encryption algorithms are standardized in 5G, specified in TS33.501:

\begin{itemize}
    \item NEA0 – plaintext with no encryption (mainly for emergency calls from devices without SIM cards) 
    \item 128-NEA1 – SNOW 3G cipher permitting backwards compatibility with 3G networks 
    \item 128-NEA2 - AES-128 CTR cipher permitting backwards compatibility with 4g-LTE network
    \item 128-NEA3 – based on the ZUC stream cipher which is specific to 5G implementations  (optional)
\end{itemize}

However, it's noteworthy in 5G specification confidentiality protection of both user and
signalling data is indicated as optional, but up to the operators to enable confidentiality in business decisions. 

In terms of security implications, the lack of confidentiality protection of user data makes data vulnerable to interception. Likewise, lack of confidentiality protection of signalling data may result in interception of status and authorization data between the UE and gNB/AMF giving opportunities for attackers to track user location and conduct other passive or active attacks.

.

\subsubsection{Integrity protection algorithms}

Four integrity protection algorithms are standardized in 5G, specified in TS33.501:

\begin{itemize}
    \item NIA0 – plaintext with no encryption (mainly for emergency calls from devices without SIM cards)
    \item 128-NIA1 – based on SNOW 3G
    \item 128-NIA2 – based on AES-128 in CMAC mode
    \item 128-NIA3 – based on 128-bit ZUC (optional)
\end{itemize}

Integrity protection is mandatory  for the signalling plane only (RRC-signalling and NAS-signalling), whilst it is optional for the user plane.

\subsection{ Protection of RAN Interfaces}

\subsubsection{CU-DU interfaces}
5G brings the concept of a split or disintegrated RAN, where gNB is separated into: Distributed Units (DU) and Central Units (CU). Communication between DU and CU is established using the F1 interface. Moreover, CUs communicate with each other via E1 interface. 3GPP specifies the use of IPsec to provide protection of the traffic over E1 and F1 interfaces, which includes the authentication, confidentiality and integrity protection. 

It's noteworthy, there is a protection requirements difference between control and user plane data: It's mandatory for the control plane interface(F1-C) and E1-C/U but optional for the user plane (F1-U) 

\subsubsection{N2 and N3 interfaces}
Interfaces N2 and N3 are interfaces that connect 5G-AN with AMF (Access and Mobility Function) and with the UPF (User Plane Function), respectively. They both carry sensitive signalling and user plane data between the access network and the core, 5G employs the IPsec and DTLS (RFC 6083) to support the certificate-based authentication and confidentiality/integrity protection.

\subsection{SBA protection}

The 5G core network also added new features to support service-driven functions, it's called SBA (Service Based Architecture). 
SBA consists of numbers of network entity functionalities, which transmit their functionality within Service Based Interfaces (SBI) that implements RESTful APIs over HTTP. The protection mechanisms of the SBA are specified in TS 33.501, which relies on TLS 1.2 and 1.3 to support authentication, and confidentiality/integrity protection of the SBI messages. 

It's noteworthy, that if the SBI interfaces are within trusted domain, then security protection is optional. 

SBA also optionally utilizes the OAuth 2.0 to support the authorization functions from different NFs(Network Functions).

\subsection{Security challenges in 5G}
There are a variety of studies on the security and privacy vulnerabilities in 5G,  NIST has recently published a special publication 1800-33B \cite{b18} which collected a list of potential threats, and vulnerabilities in 5G. Combined with many other references and reports on the 5G security and privacy, as well as some incidents reports, we summarize the major threats and vulnerabilities with 3GPP 5G in table 1.

\begin{table*}[hbt!]
\caption{Types of 3GPP 5G Threats Vulnerabilities}
\centering
\begin{tabular}{p{0.25\linewidth}p{0.25\linewidth}p{0.25\linewidth}}
\hline
 Types of Threat/Vulnerability & Sub-types & Description\\
\hline
Hardware-Based Platform & Malicious Firmware and Modded BIOS & Hardware boot up can be tampered with embedded virus or crypto-virus which leads to backdoor penetration \cite{b29,b38}\\
Remote Platform Attestation & Supply Chain Vulnerability & The hardware and software in the cloud are from various sources of vendors, some vendors may have backdoor and monitoring modules that eavesdropping the data \cite{b42}\\
NF Orchestration Enforcement & Tampered instantiated firmware & Lack of trust orchestration results in wrongfully loading of the software into weaker processor and server \cite{b29} \\
Network Function Image Encryption & trusted image misplacement &  The secure workload configuration image wasn't put in secured container but in public repository \cite{b38} \\
Infrastructure Security Monitoring & Insider attacks & lack of boundary control by deploying IDS/IPS results in inside attacks 
\cite{b40}\\
Subscriber Privacy & Exposed IMSI/SUPI & During roaming, or in an interworking domain, the IMSI/SUPI could be caught by attacker \cite{b10}  \\
Reallocation of Temporary ID & GUTI vulnerabilities & GUTI can be caught during authentication during roaming, or leak of location information \cite{b17}\\
SUCI Security & SUCI vulnerabilities& attacker may launch MITM attack during AKA handshake to request recalculation of SUCI and leads a variety of attacks \cite{b43}\\
Initial NAS Message Security & Radio bearer protection & Initial NAS messages are in clear text due to GTP vulnerabilities \\
User Plane Integrity Protection & F1-U optional protection requirement & F1-U is not required to be protected in the TS 33.501, result in the altered message payload and redirect DNS requests for DNS spoofing attack \cite{b19} \\
Cryptographic primitive selection & Downgrade attack & Some rogue operator can negotiate with the AUSF for null authentication, result in downgrade attacks \cite{b44}\\
EAP-AKA support & Non-3GPP access authentication & AKA and EAP-AKA over administrative domain result in exposure of credentials or permanent ID \\
Secure Storage & USIM credential storage & SIM card and e-SIM has some weaker protection in storage result in reading into the permanent key \\
SEAF Roaming Security & Roaming vulnerabilities & Attacker uses untrusted network to fraudulently authorize UEs \\
API Security for Network Exposure Function(NEF) & API security & N33 interface may expose user data, like SUPI \\
OAuth2 Security & Authentication Token Vulnerabilities & OAuth2 may issue token towards untrusted NF that result in unauthorized SBA access \\

\hline
\end{tabular}
\end{table*}

\section{O-RAN} 
In this section, we discuss another variant of 5G architecture, the Open Radio Access Network (O-RAN). O-RAN is a concept based on interoperability and standardization of RAN elements including a unified interconnection standard for white-box hardware and open-source software elements from different vendors \cite{b12,b25}. O-RAN architecture integrates a modular base station software stack on COTS hardware which allows base band and radio unit components from discrete suppliers to operate seamlessly together.

\begin{figure}[ht]
    \centering
    \includegraphics[width=1\columnwidth]{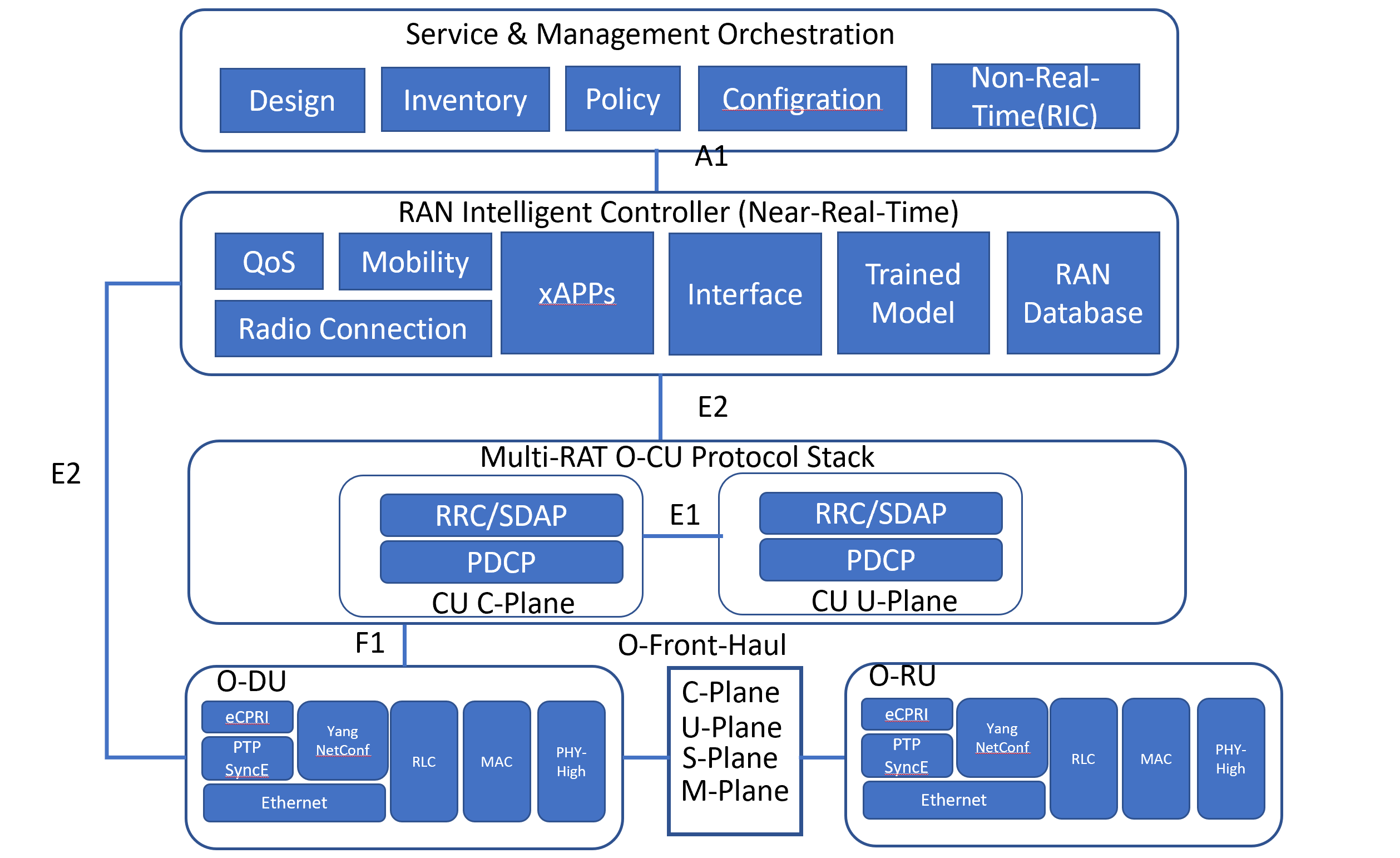}\hspace{2pc}

    \caption{O-RAN Reference Architecture}
    \label{fig:my_label}
\end{figure}

\subsection{O-RAN Security Architecture}

O-RAN benefits from the disaggregated nature of CU-DU in RAN architecture,  which inherently ensures the security, robustness and agility. On the other hand, the disaggregated, open and non-exclusive O-RAN architecture imposes more challenges for its security and privacy protection operation.  In \cite{b1}, a few telecom operators, including Deutsche Telekom, Orange, Telefónica, TIM and Vodafone, had jointly published a report on "Open RAN Security White Paper", which provides details on the Open RAN Security Focus Group (SFG) activities, and focuses on developing the four security specifications that are the pillars of the Open RAN security architecture. Figure 5 provides an overview of the O-RAN threat modeling, security requirements, protocols and tests.

\begin{figure}[ht]
    \centering
    \includegraphics[width=1\columnwidth]{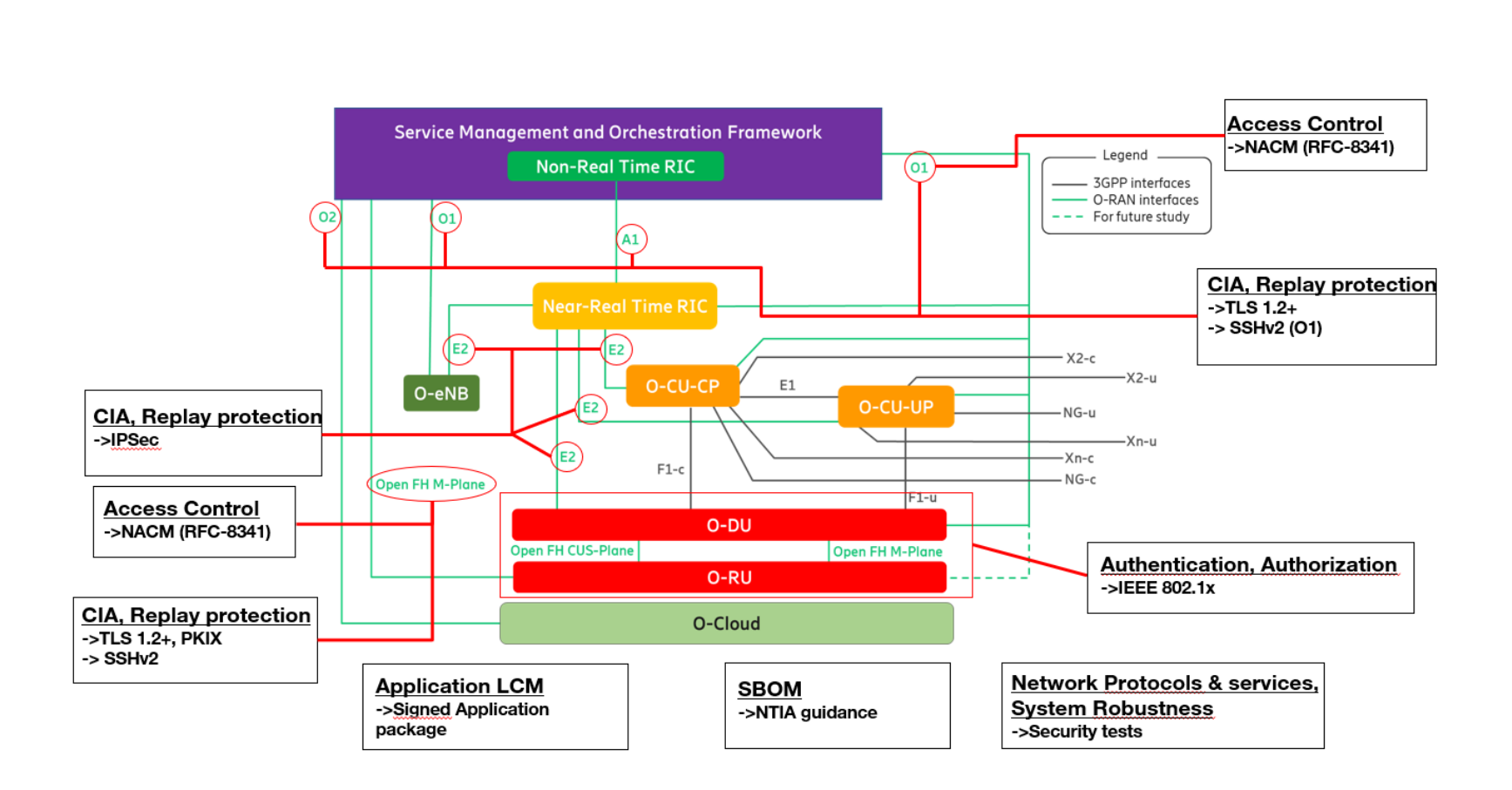}\hspace{2pc}

    \caption{View of SFG specified security requirements – November 2021– O-RAN Logical architecture \cite{b1}}
    \label{fig:my_label}
\end{figure}

O-RAN has included a suite of security requirements for O-RAN architecture in O-RAN Security Requirements Specifications v2.0 \cite{b20,b21}. The specifications mandate the interfaces with confidentiality, integrity, and availability (CIA) protection model over all the critical interfaces.
\begin{itemize}
    \item Confidentiality, Integrity, Replay protection and Data origin authentication mandatory 
requirements for A1, O1, O2 and E2 interfaces.
   \item Least Privilege Access Control on O1 interface enforcement with IETF RFC- 8341 
Network Configuration Access Control Model (NACM) requirements. 
   \item Authentication and Authorization based on IEEE 802.1x Port-Based Network Access 
Control requirements to control network access in point-to-point LAN segments across 
the Open Fronthaul interface.
   \item Mandatory support for TLS 1.2+ and Public Key Infrastructure X. 509 (PKIX) for mutual 
authentication on the Fronthaul M-Plane.
   \item Transversal requirements and tests cases for Networks Protocols and Services, DDoS attack protection, password protection policies and vulnerability scanning.
  \item  Software supply chain security support in the form of Software Bill Of Material (SBOM) 
requirements for every O-RAN software delivery following NTIA guidance. 
\end{itemize}

\subsubsection{Interface Security}

O-RAN primarily specifies the fronthaul interfaces which connects the O-DU and the O-RU.  In addition to existing 3GPP fronthaul interfaces, O-RAN defines a few new interfaces to support the disaggregated O-RAN infrastructure. 

\begin{itemize}
    \item E2 interface: The E2 interface is an open interface connecting entities, such as near-RT RIC,  Multi-RAT O-CU, or O-DU. These entities are also known as E2 entities, The E2 interface defines  procedures and functionalities for near-RT RIC to control the O-CU and O-DU. E2 also enables the performance metrics collection from O-CU and O-DU to gauge the network performance. 
    \item O1 interface: For O-RAN service operation and maintenance, O-RAN specifies the O1 interface connecting SMO and the non-RT RIC and the subordinate O-CUs and O-DUs. The O1 interface supports Management Services which includes the management of the life-cycle of O-RAN. The management life cycle consists of configuration, fault tolerance and heartbeat service. 
    \item A1 interface: The A1 interface connects specifically the non-RT RIC and the near-RT RIC, which allows the non-RT RIC to deploy policy-based guidance to the near-RT RIC. The A1 interface utilizes the JSON (Javascript Object Notation) to define the message syntax and the messages are transported over the secured HTTP channels (HTTPS).  The A1 interface relies on the A1AP protocol to policy deployment, as specified in the 3GPP framework. 
    
\end{itemize}

The following interfaces are specified in 3GPP, but are also worth mentioning here.

\begin{itemize}
    \item F1 interface: In O-RAN , The F1 interface connects a O-CU to a O-DU. This interface is applicable to the CU-DU Split gNB architecture. F1 interfaces supports the split of the control plane and the user plane control.  the F1-C control plane allows signaling between the CU and DU, while the F1-U user plane allows the transfer of application data. The F1 defines the operations and procedures in establishing the connections between the split gNB nodes (O-CU and O-DU).
    \item X1/Xn interface: The X1 interface used between RANs in LTE is reused between RAN nodes in non-standalone operation, and the Xn interface is newly specified between RAN nodes in standalone operation. In O-RAN,the extensions of X1/Xn interface further enhances the  flow control for split bearers for non-standalone operation. 
    \item E1 interface: The E1 interface manages C-plane and U-plane split on O-CU, which is to facilitate the  functional separation between different vendors. The E1 interface defines the procedures and operations that helps the interface setup, reset/configuration update and error indication, it also helps the bearer management for UEs. 
\end{itemize}

Many of the interfaces listed above rely on the underlying transport layer to obtain the level of security (confidentiality, integrity and authenticity) protection as specified in the requirements.  Similar to 3GPP, O-RAN also identifies the appropriate cryptographic primitives for each security protocol stack as table 2 summarizes. 

\begin{table*}[hbt!]
\caption{O-RAN Security Protocols}
\centering
\begin{tabular}{p{0.2\linewidth}p{0.2\linewidth}p{0.2\linewidth}p{0.2\linewidth}}
\hline
Security Protocol & Functions & Algorithms & Applicable Interfaces \\
\hline

SSH v2 & Encryption Algorithms & aes128/256-gcm  & All\\
 & & aes128/192/256-ctr \\
 & Key Agreement & ecdsa-sha2-nistp256  \\
 & & ecdsa-sha2-nistp384/521 \\
 & & ssh-ed448/25519  \\
 & Key exchange & ecdh-sha2-nistp256/384/521 \\
 & & diffie-hellman-group-exchange-sha256 \\
 & & curve25519-sha256\\
 & Message Authentication Codes & hmac-sha2-256/512-etm \\ 
 & & hmac-sha2-256/512\\
 & &umac-128 \\
TLS 1.2/1.3 & Intermediate Ciphers &tls\_ecdh\_eecdsa\_with\_aes\_128\_gcm\_sha256 \\
& & tls\_ecdhe\_rsa\_with\_aes\_128\_gcm\_sha256 \\
& & tls\_ecdhe\_ecdsa\_with\_aes\_256\_gcm\_sha384 \\
& & tls\_ecdhe\_ecdsa\_with\_chacha20\_poly1305\_sha256 \\
& & tls\_ecdhe\_rsa\_with\_chacha20\_poly1305\_sha256\\
& & tls\_dhe\_rsa\_with\_aes\_128/256\_gcm\_sha256/384\\
& & tls\_dhe\_psk\_with\_aes\_128/256\_gcm\_sha256/384 \\
& & tls\_aes\_128/256\_gcm\_sha256/384(TLS 1.3) \\
& & tls\_chacha20\_poly1305\_sha256(TLS 1.3)\\
IPsec & Supported Ciphers & IKE \\
& & ESP Tunnel Mode \\
& & X.509V3 certificate\\
& & PSK \\
DTLS & note: based on TLS 1.2+ & \\
NETCONF & note: based on TLS 1.2+ &\\

\hline
\end{tabular}
\end{table*}

\subsubsection{RIC Security}
The near-RT(Real Time) RIC is deployed at the edge of the network and
operates control loops with a periodicity between 10 ms and
1s.  The near-RT RIC consists of multiple applications, known as xApps, and of the services that are required to support the execution of the xApps.

The non-real-time (or non-RT) RIC is a component of the Service Management and Orchestration (SMO) framework. The non-RT(Real Time) RIC is one of the core components of the O-RAN architecture. Similarly to the near-RT RIC, it enables closed-loop control of the RAN with timescales larger than 1s. Moreover, it also supports the execution of third-party applications, i.e., the rApps, which are used to provide value added services to support and facilitate RAN optimization and operations, including policy guidance, enrichment information,configuration management and data analytics.

The security goal of the Non-RT RIC is to mitigate malicious xAPPs and rAPPs from leaking sensitive RAN data or from affecting the RAN performance. Both near-RT RIC and non-RT RIC are built upon multiple components running as microservices on a Kubernetes cluster. Some of the best practices are considered for RIC protection schemes:

\begin{itemize}
    \item Removing extraneous applications, libraries, and other components of the operating system in order to minimize the attack surface. Provisioning nodes with minimalist Linux distributions is a best practice.
    \item Eliminating unnecessary user accounts.
    \item Ensuring that nothing runs as root unless strictly necessary.
    \item Collecting and analyzing OS logs to detect possible breaches.
    \item Deploying OS-hardening frameworks.
    \item Define RBAC policies to  manage access to pods by users and services within the cluster.
    \item Define network policies to isolate microservice at the network level.
    \item Define admission controllers to enforce additional rules. 
    
\end{itemize}

\subsubsection{O-Cloud Security}
O-Cloud is an integral part of O-RAN which boasts the cloud native 5G RAN deployment. There are great benefits that the O-Cloud and the deployment of the RIC services, including both near-RT RIC and non-RT RIC, could become more flexible and cost-effective, and equipped with the ability to exploit CI-CD methods to deliver software patches, isolate the threats with the container method, to live migrate systems with virtualisation methods, and to roll up upgrades system to be more robust over time.

Nonetheless, the inclusion O-Cloud may bring more security challenges to the O-RAN architecture. Depending on the types of cloud services provided by private or public operators, the O-Cloud could be manifested in different forms of architecture, such as:
\begin{itemize}
    \item Private Cloud: A private cloud gives a single Operator exclusive access to and usage of the cloud service and related infrastructure and computational resources.s. It may be managed either by the Operator or by a third party Cloud Provider and may be  hosted on the Operator’s premises, known as on-site private cloud or outsourced to hosting company.
    \item Community Cloud:A community cloud serves a group of Operators that have shared concerns such as mission objectives, security, privacy and compliance policy, rather than serving a single Operator (e.g., a private cloud)
    \item Public Cloud:A public cloud is one in which the cloud infrastructure and computing resources are made available to the general public over a public network. A public cloud is owned by an organization providing cloud services, and serves a diverse pool of clients (e.g. Operators, Third parties service providers).
    \item Hybrid Cloud: A hybrid cloud is a composition of two or more clouds that remain as distinct entities but are bound together by standardized or proprietary technology that enables data and application portability.
\end{itemize}
Many forms of the cloud deployment types have implicated the security design, some of the security threats and vulnerabilities are captured in section 3.1.4.

\subsubsection{Challenges for O-RAN security}
In this section, we summarize the potential security threats and vulnerabilities within each sub system. Due to the complex nature of virtualization and cloud technologies, it's almost impossible to capture all attacking methods and its attacking surfaces exhaustively. We can utilize the listed threats and vulnerabilities in table 3 to systematically categorize the O-RAN security and vulnerabilities, and help to identify more derivative attacks in the future. It's worth mentioning that O-RAN also inherits most of the threats and vulnerabilities listed in table 2 due to the fact O-RAN is still based on 3GPP architecture with cloudified architecture. That being said, O-RAN has more attacking surfaces than its counterpart in 3GPP 5G. 
\begin{table*}[hbt!]
\caption{Types of O-RAN Threats and Vulnerabilities}
\centering
\begin{tabular}{p{0.25\linewidth}p{0.25\linewidth}p{0.25\linewidth}}
\hline
 Types of Threat/Vulnerability & Sub-types & Description\\
\hline
O-Cloud and RIC & Software flaw attack &  Opensource software, opensource libraries, 3rd party components may allow attacker to exploit O-Cloud environment \cite{b26,b27,b30} \\
& &  -Compromise of the underlying VM, Container \cite{b37}\\
& &  -Exploit host access via Escape to Host \\
& &  -Take advantage of weak identity and access management policies to attempt to elevate privileges\\
& &  -Execute adversary-controlled code \cite{b27,b29} \\
& &  -Enable adversary to move from a virtualization environment onto the underlying host \cite{b37}\\
& Malicious access & Adversaries may obtain and abuse credentials of existing accounts as a means of gaining initial access, persistence, privilege escalation, or defense evasion \cite{b40}\\
& Untrust fusion of VM & vulnerability that purposely run services to different VMs or Hypervisor engine \cite{b39}\\
& Weak Authentication and Authorization & Lack of strong authentication and authorization between untrusted entities \\ 
& VM/Container Data Theft & Sensitive data insecure storage or malware expose the data \cite{b38}\\
& VM migration attack & VM/ container system vulnerabilities may lead to attacks on signaling and data \cite{b38}\\
& Unauthorized change resource allocation & compromised IMS / DMS (Infrastructure Management Service) / (Deployment Management Service) components can change the resources allocation for targeted service \\
& VM/Container image tampering & Malicious code injection to tamper the VM or containers image repository \cite{b38}\\
& VM/Container hyperjack & The VM/Container running on weak host machine which can be manipulated by adversary  \\
& Server Boot Tampering & Cloud server without TPM \cite{b38}\\
& Cloud inside attacks & Malware or less enforced access policy allows attacks from inside \cite{b40} \\
\hline
Interface Security & O2 interface vulnerabilities & Man-in-the-Middle attack on O2 interface as O2 interface is not protected \\
& Attack on interfaces between RIC and the virtualization layer & Man-in-the-Middle attacks on open interfaces \cite{b12} \\
& & DoS/DDoS attacks on all interfaces \cite{b3}\\
&Weak authentication protocol & Downgrade attack on Authentication and authorization protocol with weak security primitives\\
& Air interface signaling attacks & GTP-U protocol vulnerabilities expose user ID and location information \cite{b41}\\

\hline

\end{tabular}
\end{table*}

Similar to RIC security best practices, there is a plethora of useful and practical guidance to mitigate the attacks listed above, such as Docker best practices, MITRE container matrix, CIS VMWARE Benchmark,  CISA/NSA Kubernetes security guidance etc. 

O-RAN architecture has also opened up the space allowing to plug in AI/ML models to automatically and efficiently manage network resources in various use cases such as traffic steering, quality of experience prediction, and anomaly detection. Unfortunately, some of the researches \cite{b46} have demonstrated that the AI/ML models could become vulnerable in different contexts of attacks or could be utilized to provide extra attacking surface for breaking the security and privacy of O-RAN. We listed a few risks associated with AI/ML models in table 4.

\begin{table*}[hbt!]
\caption{Types of AI/ML Threat Model}
\centering
\begin{tabular}{p{0.25\linewidth}p{0.25\linewidth}p{0.25\linewidth}}
\hline
 Types of Threat/Vulnerability & Sub-types & Description\\
\hline
\hline
AI/ML Threats  & Data Staining  & The adversarial inputs "stained" data into the AI/ML model to generate wrong output, i.e the radio control is no longer optimized \\
& AI/ML Model corruption &  The AI/ML model structure is compromised and generate biased output, i.e The QoS/QoE model always allocate a certain UE to a prefixed time/frequency domain or a cell \\
& Subscriber Privacy Inference & The adversarial utilizes AI/ML model to observe the traffic behavior to reveal subscriber's privacy information, i.e user's location information \\
& Data Reconstruction &  The adversarial causes the AI/ML model to leak information to obtain the training date set which is crucial for AI/ML model \\
& Model Extraction & The adversarial attempts to extract information about the model by polling the network in order to train a replica of the model, i.e replicating the model used for QoS/QoE classification. \\
& Resource Exhaustion & The adversarial causes the AI/ML model to use more resources \\
\hline
\end{tabular}
\end{table*}
\section{5G with Zero Trust Architecture}

\subsection{Zero Trust in standards}
The work of the Jericho Forum in 2004 publicized the idea of deperimeterization—limiting implicit trust based on network location and the limitations of relying on single, static defenses over a large network segment \cite{b7}, which can be regarded as the incubation of the zero trust concept.  NIST has published SP 800-207 encompassing a set of ZTA design guidelines for various types of infrastructure.   The Defense Information Systems Agency (DISA) and the Department of Defense also published their work on a more secure enterprise strategy dubbed “black core”\cite{b33}. 

 There are initiatives from US federal agencies which builds secrecy capabilities and policies, such as the Federal Information Security  Modernization Act (FISMA) followed by the Risk Management Framework (RMF), Federal 
Identity, Credential, and Access Management (FICAM), Trusted Internet Connections (TIC), and Continuous Diagnostics and Mitigation (CDM) programs, all push for adoption of  zero trust model federal agencies. A clear definition of the zero trust comes from NSA Zero Trust Architecture (RA), in which Zero Trust is defined as:

\begin{quote}
 \textit{Zero Trust is a cybersecurity strategy that embeds security throughout the architecture for the 
purpose of stopping or mitigating data breaches and reducing cybersecurity operational risk. 
This data-centric security model eliminates the idea of trusted or untrusted networks, devices, 
personas, or processes and shifts to multi-attribute-based confidence levels that enable 
authentication and authorization policies under the concept of least privileged access.}
\end{quote}

Industries such as Google, also released its own Zero Trust Architecture, known as Beyondcorp \cite{b8}. Beyondcorp is an internal Google initiative to enable every employee to work from untrusted networks without the use of a VPN. Now, BeyondCorp is used by most Googlers every day to provide user- and device-based authentication and authorization for Google's core infrastructure and corporate resources.

\subsection{5G Zero Trust Reference Architecture}

3GPP and O-RAN also work on zero trust architecture. From an architectural point of view, it's required to remove implicit security perimeter assumptions that existed long before 5G. For example,  the packet core is normally considered as a trusted domain. 3GPP had developed an entity known as SEPP to enforce the perimeter protection at the security perimeter of the packet core, which indeed reflects on the design philosophy - "a centralized trust model". Nonetheless the centralized trust model imposes itself as a challenge towards the cloudified O-RAN 5G architecture , as the major security attacks and harms are from the "inside" (of the perimeter).  We recognize that aforementioned 5G and O-RAN risks makes Zero Trust an appealing protection option so as to manage the risks and threats, and reduce the attacking surfaces and environments.  We propose an overlay 5G zero trust reference model in Figure 6. 

\begin{figure}[ht]
    \centering
    \includegraphics[width=1\columnwidth]{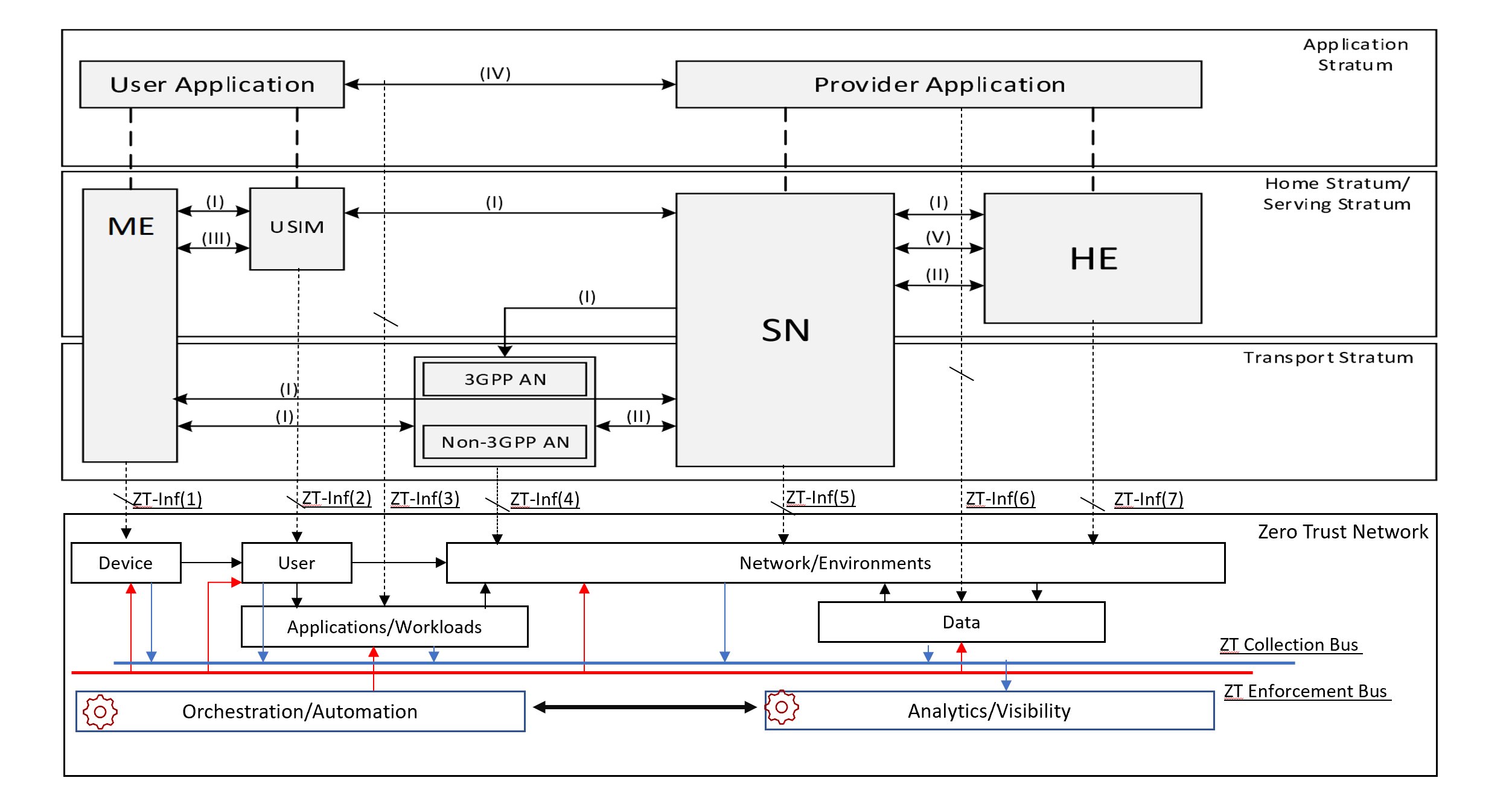}\hspace{2pc}

    \caption{5G Zero Trust Reference Model}
    \label{fig:my_label}
\end{figure}
 
The 5G Zero Trust reference model is an overlay model,  5G architecture runs seamlessly atop the underlying Zero Trust Network providing the Zero Trust services, within which the ZT services are povided by 7 ZT classes. The main goal of the 5G zero trust model is to preserve the integrity of 5G Architecture while obtaining the Zero Trust capabilities, in the meantime, reducing the complexity and overhead of the application of ZT services. Many of the zero trust principles, such as I) Never Trust Always Verify, II) Assume Breach and III) Verify Explicitly,  are being embedded in the Zero Trust Network. The zero trust services are provided in different classes in table 5.

\begin{table*}[hbt!]
\caption{5G Zero Trust Reference Model}
\centering
\begin{tabular}{p{0.25\linewidth}p{0.25\linewidth}p{0.25\linewidth}}
\hline
 ZT Types & ZT Services\\
\hline
Data Class & Privacy Protection Service \\
& Digital Right Management Service \\
& Data Privacy Preserving Service \\
& Data Labeling \\
\hline
Users Class & Authentication ZT services \\
& Multi Factor Authentication \\
& Access Control Privilege  \\
\hline
Devices Class & Signalling Protection  \\
& Configuration Management \\
&Antivirus, Anti-Malware \\
&Authorization\\
\hline 
Networks/Environments Class & Segmentation \\
& Network Attestation \\
& Authenticated Encryption service \\
& Routing Protection \\

\hline
Application/Workloads Class & SecDevOps \\
& Virtualization and Containerization Protection Service\\ 
& Secure Token distribution service \\
\hline
Orchestration/Automation Class & Threat response \\
& Policy Enforcement \\
& Configuration Management \\
& Process Automation \\
\hline
Analytics/Vistiblity  Class &  Security information and event management(SIEM) \\
& Threat Detection \\
&Event Logging \\
& Unified Access Management (UAM) \\
\hline

\end{tabular}
\end{table*}

There are 2 classes that powers the 5 ZT service classes: I) Orchestration/Automation, II) Analytics/Visibility, they interact with each other in order to i)collect and analyze data \& events, ii) make policy decision, and iii) enforcing these policies. 

In the following subsections, we will look into a few potential areas suitable for ZT Network, that can provide enhanced zero trust service in corresponding class of 5G Zero Trust Reference Model. 

\subsubsection{Security in radio interface -- Device \& Network Class}
Radio interface is normally overlooked for security even though in 3GPP TS 33.501 the signaling messages protection are specified as mandatory. Encryption are provided by GPRS Tunneling Protocol (GTP) which has been criticized for weakness \cite{b34}. 

One potential weakness in 5G is the keys exposure for eavesdropping or hijacking. For instance  the keys for radio interface encryption are obtained in the home core network and then sent to the visiting radio network using signaling channels during UE roaming. When sent between network nodes, the connection between operators’ signaling systems should be adequately secured so that the radio interface session keys can be transferred via SS7 and DIAMETER, and such exposure is prevented. 

Another potential threat on the radio interface is the signaling storms. Massive IoT deployment have several limitations such as computational capabilities, energy support, and memory capabilities. At the same time, these low-cost devices can be compromised and allow DoS and DDoS attacks against the radio access network. Unexpected non-malicious events may also cause the devices to behave abnormally and produce “flash crowd” \cite{b11} situations, leading to the exhaustion of radio resources.

The signalling messages should be protected with more efficient mutual authentication on both UE and RAN, further enhanced protection may employ encryption/decryption mechanisms to obtain full secrecy.

\subsubsection{Context based authentication and authorization -- Data \& User Class}
Adoption of the zero-trust principle demands all messages exchanged over the air interface are authenticated, authorized, and continuously validated to be granted access to resources. The authentication today is still reliant on DIAMETER protocol which lacks the capabilities to control the access privileges based on context. Some of the attacks may utilize the embedded or piggybacked messages \cite{b35} to gain unauthorized access to resources. 

DPI (Deep Packet Inspection) provides some levels of context based access control together with policy server (PA/PEP combo). We envision more proactive access control mechanisms by utilizing AI/ML to adapt to new attacking vectors would be able to achieve high efficiency in ZTA.

\subsubsection{Network Attestation -- Analytics/Visibility Class}
Attestations stems from  the industry efforts around the Trusted Platform Module (TPM hardware), whereby the trust attestation is towards the TPM embedded system, i.e a server or a PC.  The threat of hardware vulnerability motivated the computing industry to form the Trusted Computing Group (TCG) where the notion of a hardware root-of-trust was used to distinguish the security relevant portions of a hardware platform. The TCG defined trusted computing more organically by building upon granular components that were described as shielded locations and protected capabilities. Shielded locations are defined as:
\begin{quote}
\textit{" A place (memory, register, etc.) where it is safe to operate on sensitive data; data locations that can be accessed only by protected capabilities"}
\end{quote}

Protected capabilities are defined as:
\begin{quote}
\textit{"the set of commands with exclusive permission to access shielded locations"}
\end{quote}

By extension, all components that could be classified as shielded locations or protected capabilities is what defines the hardware security domain.

We can extend the concept of the attestation to the network level by utilizing power of the blockchain \cite{b36}. Network attestation addresses the threats from the notorious inside attacking as it prohibits the \textit{"untrustworthy"} node from participating the network communication.

\subsubsection{Proximity based PDP/PEP -- Orchestration/Automation Class}
In NIST SP 800-207, it outlines the ZTA design basics in chapter 2, in which the policy servers PDP/PEP (Policy Decision Point and Policy Enforcement Point) are the essential components of a Zero Trust Architecture. The PDP/PEP  passes proper judgment to allow the subject to access the resource. PDP/PAP conducts the basic authentication and authorization of an entity. The PDP/PEP applies a set of controls so that all traffic beyond the PEP has a common level of trust. The PDP/PEP cannot apply additional policies beyond its location in the flow of traffic. To allow the PDP/PEP to be as specific as possible, the implicit trust zone must be as small as possible.  The implicit trust zone ensures the robustness of the whole system in the events of attacks. This brings up a question: how efficient is it to manage the implicit trust zones at atomic level in the magnitude of tens of thousands of IIoT 5G deployment?  The answer may be found in section 4.2. 

We argue that the proximity of the PDP/PEP decides the responsiveness and robustness of the ZTA. A far-fetched PDP/PEP located at remote office/headend may not provide the accurate policies towards the events ( join, leave, and other disruptive activities) and be real time. We envision the clustered PDP/PEP design at the proximity of the targeted implicit trust zones would help to reduce the stress of the policy management for thousands of autonomous IIoT networks in 5G.

\subsubsection{AI/ML assisting threat repository and prediction -- Analytics/Visibility Class}
The cloudified 5G network attacking surface is massive, and continuing to be exploited with new tools and means. There would be several hundred billion signals that need to be analyzed to accurately evaluate the risks and identify the potential attacks. Analyzing attacks and improving the trustworthiness need revolution, in direct comparison with human powered cybersecurity analysis methodologies. A successful ZTA deployment doesn't mean 100\% proof of attacks, but it means early detection of attacks and system prediction to ensure the PDP/PEP accurately update and manage the policies to reflect the new vectors of attacks, attestation network is able to pick up hidden mischief with higher level of confidence.

We envision that Artificial Intelligence (AI) and Machine Learning (ML) based tools for cybersecurity will be able to help 5G networks reduce breach risk and improve the security assurance efficiently and effectively.  From malware exploiting zero-day vulnerabilities to identifying risky behavior that might lead to a phishing attack or download of malicious code, AI/ML extends the IT security specialist's capability to identify the potential attacks quickly and efficiently, it also helps to predict the new attacking surfaces in the events of network topology changes, new nodes being added, new signaling transmitted.  

\subsection{Cost effectiveness of ZTA in 5G}

It's not effortless to apply the zero trust strategy over 5G as the principles indeed requires enormous overhead in achieving the ultimate security goals, especially the computational resources allocated for the authorization, authentication and  verification operations attached to each message. Some  studies model the cost of applying ZTA towards enterprise networks be proportional to the number of employees, Table 6 excerpted from \cite{b2} provide a cost estimate towards the ZTA in different types of enterprise networks. By using this model, we can roughly estimate a mid-sized operator's 5G network with 1 million subscribers. 

\begin{table*}[th!]
\caption{ZTA Tools and Current Price as of 2022}

\centering
\resizebox{\textwidth}{!}{\begin{tabular}{|l|l|l|lllll}
\hline
 &
   &
   &
  \multicolumn{5}{l|}{Cost Based on the Number of Employees in an Organization} \\ \cline{4-8} 
\multirow{-2}{*}{Tools and Resources} &
  \multirow{-2}{*}{ZTA Component} &
  \multirow{-2}{*}{Unit Cost per Year} &
  \multicolumn{1}{l|}{1-100} &
  \multicolumn{1}{l|}{101-400} &
  \multicolumn{1}{l|}{401-700} &
  \multicolumn{1}{l|}{701-999} &
 \multicolumn{1}{l|}{1,000,000} \\ \hline
CylanderPROTECT for Endpoint Protection(Cylance, 2022) &
  Endpoint Protection -Subject and Resources &
  \begin{tabular}[c]{@{}l@{}}\$45 per endpoint for 1-250 endpoints.\\ \$41.75 per endpoint for 501-1000 endpoints\end{tabular} &
  \multicolumn{1}{l|}{\$4,500.00} &
  \multicolumn{1}{l|}{\$16,700.00} &
  \multicolumn{1}{l|}{\$29,225.00} &
  \multicolumn{1}{l|}{\$41,708.25} &
  \multicolumn{1}{l|}{} \\ \hline
\begin{tabular}[c]{@{}l@{}}Kaspersky Security for Business\\  with encryption service (G2,2021a)\end{tabular} &
  \begin{tabular}[c]{@{}l@{}}Data Encryption- Subject, Resources, and PEP\\ Endpoint Protection -Subject and Resources\end{tabular} &
  \$45 per node &
  \multicolumn{1}{l|}{\$4,500.00} &
  \multicolumn{1}{l|}{\$18,000.00} &
  \multicolumn{1}{l|}{\$31,500.00} &
  \multicolumn{1}{l|}{\$44.955.00} &
  \multicolumn{1}{l|}{} \\ \hline
\begin{tabular}[c]{@{}l@{}}Google Workspace for cloud-based \\ Storage and Access Control (Prokopets, 2022)\end{tabular} &
  \begin{tabular}[c]{@{}l@{}}Access Control -PE in PDP\\ Data Encryption -Subject, Resources, and PEP\\ Cloud Storage -Resource\end{tabular} &
  2 TB pooled cloud storage for \$96 per user &
  \multicolumn{1}{l|}{\$9,600.00} &
  \multicolumn{1}{l|}{\$38,400.00} &
  \multicolumn{1}{l|}{\$67,200.00} &
  \multicolumn{1}{l|}{\$95,904.00} &
  \multicolumn{1}{l|}{} \\ \hline
Microsoft OneDrive (Prokopets, 2022) &
  \begin{tabular}[c]{@{}l@{}}Access Control - PE in PDP\\ Data Encryption -Subject, Resources, and PEP\\ Cloud Storage -Resources\end{tabular} &
  1 TB for \$60 per user &
  \multicolumn{1}{l|}{\$6,000.00} &
  \multicolumn{1}{l|}{\$24,000.00} &
  \multicolumn{1}{l|}{\$42,000.00} &
  \multicolumn{1}{l|}{\$59,940.00} &
  \multicolumn{1}{l|}{} \\ \hline
Virtu Encryption (G2, 2021c) &
  Subject, Resources, and PEP &
  \$949 per 4 user &
  \multicolumn{1}{l|}{\$18,960.00} &
  \multicolumn{1}{l|}{\$75,840.00} &
  \multicolumn{1}{l|}{\$132,720.00} &
  \multicolumn{1}{l|}{\$189,410.40} &
  \multicolumn{1}{l|}{} \\ \hline
\begin{tabular}[c]{@{}l@{}}Microsoft Azure Active Directory\\ for Access Control (Zelleke, 2021)\end{tabular} &
  PE in PDP &
\begin{tabular}[c]{@{}l@{}}Comes free with Microsoft OneDrive or \\ \$72 per user seperately\end{tabular} &
  \multicolumn{1}{l|}{\$7,200.00} &
  \multicolumn{1}{l|}{\$28,800.00} &
  \multicolumn{1}{l|}{\$50,400.00} &
  \multicolumn{1}{l|}{\$71,928.00} &
  \multicolumn{1}{l|}{} \\ \hline
 {\cellcolor{blue!25}} \begin{tabular}[c]{@{}l@{}}Estimate for Mid-sized 5G Operator with 1 million\\ Subscribers \\ (Assuming based on Google Workspace)\end{tabular} &
 {\cellcolor{blue!25}} \begin{tabular}[c]{@{}l@{}}Access Control -PE in PDP\\ Data Encryption -Subject, Resources, and PEP\\ Cloud Storage-Resource\end{tabular} &
 {\cellcolor{blue!25}}   2 TB pooled cloud storage for \$96 per user &
  \multicolumn{1}{l|}{ {\cellcolor{blue!25}}  } &
  \multicolumn{1}{l|}{ {\cellcolor{blue!25}}  } &
  \multicolumn{1}{l|}{ {\cellcolor{blue!25}}  } &
  \multicolumn{1}{l|}{ {\cellcolor{blue!25}}  } &
  \multicolumn{1}{l|}{ {\cellcolor{blue!25}}  $\sim$\$96,000,000.00} \\ \hline
\end{tabular}}

\end{table*}

Table 6 showcases ZTA tools selected from a few of well known solution providers including Google, Kaspersky, and Microsoft Azure etc. At end of the table we added an estimation for a mid-sized 5G operator with 1 million subscribers, the estimation is based on the cost of Google Workspace for cloud based storage and access control. It's obvious the cost associated with ZTA is extraordinarily high with close to \$100 million for 1 million subscribers (the highlighted part within table 4). We argue that the cost is linearly calculated with the base of \# of subscribers is indeed overlooking the amount of critical messages should be processed according to ZTM, meaning a practical ZTA implementation in 5G could be higher if per message verification is implemented.  It's understood that ZTA's central components are the PDP/PEP (Policy Decision Point and Policy Enforcement Point) which act on all events and messages exchanged within the network, and every NF and component are under the control of policy servers. We speculate the main reason behind the high cost of ZT in 5G is due to the fact of large amount of resources allocated to the zero trust based verification and processing. 

Overall, it's said that implementing ZTA within 5G is very expensive. We need to carefully address the cost effectiveness issues towards ZTA in 5G air interfaces.

\section{Conclusion}
This paper is aimed to discuss the security and privacy landscape in the new era of 5G, in which virtualization and cloudification is the epicenter of of 5G architecture. We first presented a comprehensive overview of the 3GPP 5G and O-RAN security specifications, architectures and operations. We then introduced 3GPP 5G security architectural Network Function. We then presented the O-RAN 5G security architecture and related Network Functions, and new interfaces including E2, O1/O2, A1, the fronthaul interface. We  discussed the threat models associated with each 5G technology respectively.  We also deep dive into the potential integration of Zero Trust Model into 5G security architecture, and discuss how the Zero Trust design principle could benefit the security and privacy, we also cautiously pointed out the cost of implementing Zero Trust in 5G could be prohibitive.

\section{Acknowledgement}
The authors would like to thank my colleagues from Dell Technologies, Herb Kesley,  Alex Reznik, Qing Ye, and Erik Boch  for their helpful guidance and profound knowledge of 5G security technologies.

\vspace{12pt}

\end{document}